# Field-driven dynamics of magnetic Hopfions

David Raftrey[1,2] and Peter Fischer[1,2]*

[1]Materials Sciences Division, Lawrence Berkeley National Laboratory, Berkeley, CA 94720, USA

[2]Physics Department, UC Santa Cruz, Santa Cruz CA 95064, USA

*Corresponding author: PJFischer@lbl.gov

**Abstract**

We present micromagnetic simulations on resonant spin wave modes of magnetic Hopfions up to 15 GHz driven by external magnetic fields. A sharp transition is found around 32 mT coinciding with a transition from Hopfions to magnetic torons. The modes exhibit characteristic amplitudes in frequency space accompanied by unique localization patterns in real space, and are found to be robust to damping around topological features, particularly vortex lines in Hopfions and Bloch points in torons. The marked differences in spin wave spectra between Hopfions, torons and target Skyrmions can serve as fingerprints in future experimental validation studies of these novel 3d topological spin textures.

**Introduction**

Topological solitons, originally proposed to explain quantized particles in continuous fields [1-3], have recently seen significant attention in condensed matter research, where they appear as stable configurations of the magnetization field in ferromagnetic materials [4, 5]. Magnetic Skyrmions are prominent examples [6], which exist in two dimensional chiral magnets resulting from a competition between symmetric exchange (Heisenberg) interactions with antisymmetric exchange i.e., Dzyaloshinski-Moriya interaction (DMI) [7, 8]. Their topological protection makes them attractive candidates for future high-density and low-power spintronic devices [9], but their dynamical behavior is often impaired by topological Hall effects that can be attributed to a gyrovector [10, 11]. Recently, the generalization of magnetic Skyrmions into the third dimension has opened a path towards more complex and diverse topological solitons, including rings, knots and links. Recently, several new topological spin structures have been observed including target Skyrmions [12, 13], Skyrmion tubes [14], chiral bobbers [15], vortex rings [16], and Hopfions [17]. Whereas the topological charge that characterizes magnetic Skyrmions in two dimensions is the winding number $N_{sk} = \frac{1}{4\pi} \iint d^2r \boldsymbol{m} \cdot (\frac{\partial \boldsymbol{m}}{\partial x} \times \frac{\partial \boldsymbol{m}}{\partial y}))$ with $\boldsymbol{m}$ being the magnetization, the corresponding three-dimensional topological solutions can be classified by an additional topological invariant, the Hopf number ($Q_H$). In real space this can be expressed as $Q_H = -\int \boldsymbol{B} \cdot \boldsymbol{A} d^3r$ with $\boldsymbol{B}$ being the emergent magnetic field from the spin texture and $\boldsymbol{A}$ is the magnetic vector potential [18].

The topological protection in three dimensions is analogous to the process that stabilizes Skyrmions in two dimensions. Skyrmions cannot be smoothly unwound into a uniformly magnetized state without passing through a divergent singularity, so they are stable even when they are not in the lowest energy state. Magnetic Hopfions take the form of toroidal knots where each iso-surface, the area of a single direction of magnetization, is a loop. The Hopfion's topological invariant is its Hopf number $Q_H$ which is the number of times these iso-surfaces are linked (see Fig. 1a). A Hopfion can therefore be considered as a twisted, closed loop of Skyrmion string, where the Hopf number is the number of twists [19]. With the application of a static magnetic field along the z-axis, the Hopfion constricts in diameter, until at a certain field value, it transforms to the un-linked toron state. The toron is equivalent to a Skyrmion tube closed with two opposite Bloch points at the top and bottom (Fig. 1b).

Hopfions are predicted to have complex dynamics relevant for spintronic applications [20, 21]. Previous computational studies have investigated the dynamics of Hopfions under spin transfer torque [22, 23]. Because of their vanishing gyrovector, and therefore vanishing Skyrmion Hall Effect, under the rigid-body approximation Hopfions can move in a racetrack geometry without deflection, however, they still exhibit complex tumbling and breathing dynamics in three dimensions [23].

In addition to current-induced transport dynamics, spin textures can also be excited by external oscillating fields. The three-dimensional resonances of target Skyrmions have been precisely simulated [12], recent studies on Skyrmion tubes [24] showed mode-dependent spin wave propagation, and three-dimensional resonance experiments with Magnetic Field Microscopy revealed a complex spin wave spectrum of magnetic "nanovolcanos" [25].

Here we present micromagnetic simulations of the resonant dynamics of magnetic Hopfions confined in chiral magnetic nano-pillars up to 15 GHz as a function of external magnetic fields and as a function of damping parameter. We find that Hopfions have unique field-excited dynamics distinct from target Skyrmions and torons, which opens a path to discriminating between Hopfions and target Skyrmions based on their breathing mode dynamics.

**Simulation details**

To mimic an experimental scenario [17] a stable Hopfion state was simulated in a disk of chiral ferromagnetic material which is sandwiched between two layers of material with high perpendicular magnetic anisotropy (PMA) (see Fig 1). The high PMA layers enforce a uniform M = (0,0,1) boundary condition on the top and bottom of the disk. Without the uniform boundary condition, the Hopfion expands until it hits the surface and breaks its linked iso-surfaces, transforming into the related target Skyrmion state [5]. It has been shown that a range of anisotropies can stabilize a bound Hopfion. For simplicity, the simulations shown here employ a fixed spin approximation of the high PMA layer on the top and bottom of the chiral disk. The geometry and material parameters are based on previous theoretical investigations of Hopfions confined in magnetic heterostructures [19, 26].

The system was simulated using the MuMax3 GPU accelerated micromagnetic simulator package [27]. Simulations were run in parallel on the Lawrencium supercluster at Lawrence Berkeley National Laboratory in Berkeley, CA. Additional simulations were performed in OOMMF [28].

The local energy density of the chiral disk is given by

$$H = -A(\boldsymbol{\nabla} \cdot \boldsymbol{m})^2 - D\boldsymbol{m} \cdot (\boldsymbol{\nabla} \times \boldsymbol{m}) - \mu m_s \boldsymbol{H} \cdot \boldsymbol{m}$$

where the exchange constant was chosen as $A$ = 2.19 pJ/m, the DMI constant $D$ = 0.395 mJ/m$^2$, and the saturation magnetization $m_s$ = 384 kA/m. The geometrical dimensions of the chiral disk were height h = 90 nm and diameter d = 200 nm. The spins in the top and bottom 2 nm layers are fixed in the (0,0,1) direction. This specific geometry based on material parameters is needed to stabilize a Hopfion. The dimensions of the disk are set by the chiral period of the material parameters L = 4$\pi$A/D = 70 nm, which rescales the disk geometry in units of L to h= 1.28 L and d=2.86 L. These dimensions allow the magnetization to rotate by 2$\pi$ from top to bottom of the Hopfion and by 4$\pi$ across the diameter of the Hopfion. The cell size was chosen to be 2nm X 2nm X 2 nm for a total 450,000 cells per simulation. For computational efficiency, the non-local demagnetization energy is excluded.

To simulate magnetization dynamics, MuMax3 numerically integrates the Landau-Lifshitz-Gilbert equation

$$\dot{\bm{m}} = -\gamma \bm{m} \times \bm{H} + \alpha \bm{m} \times \dot{\bm{m}}$$

A range of damping parameters $\alpha$ = {0.001, 0.1} was used to explore the GHz spin wave spectrum of the system and to determine where distinct spin resonances are replaced at high damping by large displacement breathing modes. This range of damping parameters explores the robustness of the modes.

To excite spin wave resonances a pulsed magnetic field was applied. The sinc function, $H(t) = H_{max} \frac{\sin(2\pi\omega_{max}t)}{2\pi\omega_{max}t}$, was used as the pulse because its Fourier transform is a step function up to a maximum frequency $\omega_{max}$, so it excites an unbiased spin wave response from the Hopfion for frequencies up to $\omega_{max}$ (see Figs. 2b, 2c.) The sinc pulse $\omega_{max}$ frequency was 15GHz with an amplitude of 5mT. The pulse is arbitrarily offset in time to peak at 0.67ns. Low damping ($\alpha$ =0.001) spin resonance simulations were run for 20 ns with data taken every 17 ps to ensure a resolution below the Nyquist frequency 1/2$f$.

In the high damping $\alpha$ =0.1 simulations a sinusoid, $H(t) = H_{max}\sin(2\pi\omega_{max}t)$ with $\omega = 1 GHz$ and amplitude of $H_{max} = 12\ mT$ was used to excite large amplitude breathing modes. The sinusoid was chosen here instead of the pulse because the pulsed excitation quickly decays under large damping. These high damping simulations were run for 10 ns simulation time with data recorded at increments of 0.1 ns.

**Spin wave breathing mode spectrum under pulsed magnetic field**

The spin wave resonances were excited by a sinc pulse magnetic field in the +z direction. The resonances are calculated by taking the spatially average fluctuation in z magnetization $\langle \delta m_z(t) \rangle = \langle m_z(t) - m_z(0) \rangle$. The $\langle \delta m_z(t) \rangle$ signal is the Fourier transformed signal to compute the frequency of the resonances.

To examine how a static magnetic field affects the spin wave spectrum, a set of simulation was performed with an additional static magnetic field in +z in addition to a pulse excitation.

Fig. 2a shows the obtained Power Spectral Density (PSD) response of a Hopfion to the sinc pulse (see Fig. 2b) under static applied field where the simulations were performed in steps of 1 mT from 0 to 50 mT. Whereas the Hopfion is stable at zero applied field, the frequency spectrum shifts to lower frequencies under increasing magnetic field as the Hopfion constricts. At a critical field of 32 mT the Hopfion transforms into an un-linked toron state, and this topological transition is accompanied by a discontinuity in the spin wave spectrum. Although a similar number of resonances are present for the Hopfion and toron, which are labeled in Fig. 2a with h.1 – h.5 for the Hopfions and t.1 – t.5 for the torons, the spin wave spectrum is discontinuous across the transition for modes h.2-5 and t.2-5, and only mode h.1 is continuous across the transition.

The width of the resonances modulates under the changing of applied fields, which not only intensifies when approaching the critical field, but implies that Hopfions have a field dependent quality factor to their resonances.

Fig. 3a and c show the location in frequency space of the five Hopfion and toron resonances at 10 mT and 40mT, respectively. To visualize each resonance spatially, a similar process is applied cell by cell. The single cell fluctuation $\delta m_z(r,t)$ is sampled and a Fourier transform is taken for the $\delta m_z(r,t)$ signal of each cell, showing how the amplitude at each frequency is distributed spatially. In Fig. 3b and d the regions with the highest amplitude in the PSD at the resonant frequencies are highlighted in green for the out-of-plane $M_z$ component and in blue and red for the in-plane $M_x$ and $M_y$ components, respectively.

The Hopfion responds to the pulsed field with breathing resonances in $M_z$ accompanied by twisting resonances in $M_x$ and $M_y$ (Fig. 3b). The twisting in $M_x$ and $M_y$ indicates that the spin waves are of hybrid

Bloch/Néel character. The complexity of the three-dimensional spin texture makes these hybrid spin waves most energetically favorable. These modes are localized in concentric rings around the torus. The mode h.2 has the largest amplitude and is localized in a torus at the center of the Hopfion, with additional symmetric components on the upper and lower vortex ring. In these regions, the spins are mostly in-plane (Fig. 3b). The high amplitude in these regions is expected, as the interaction between magnetization and external magnetic field is strongest when the magnetization is orthogonal to the field. The second most significant mode h.3 is active only at the vortex lines (Fig. 3b). A less significant mode h.1 couples the Hopfion to the edge of the disk. This mode has the lowest frequency, and largest spatial extent. The higher frequency modes h.4 and h.5 are localized at concentric rings around the Hopfion, mostly concentrated on one or both vortex lines or one of several tori around the center of the Hopfion. An interesting feature of the modes is that they spatially overlap, with each mode localized to a different combination of the same regions (Fig. 3b). Whereas modes h.1-h.4 all involve one or both vortex lines, modes h.2 and h.5 have a component localized mostly on a central torus. This shows that the system achieves different resonant frequencies by distributing the spin wave in unique combinations of highly excitable regions.

This situation is very different for the torons, where the spin waves have highest amplitude near the Bloch points. The dominant mode, t.2, includes the Bloch point as well as the edge of the disk. Interestingly, both the toron and Hopfion have their largest excitations localized near topological defects, vortex lines for the Hopfion, and Bloch points for the toron. The Hopfion modes exhibit a certain analogy to the toron modes. Modes that occur at similar frequencies are localized in similar positions on the spin texture e.g., modes h.3 and t.3 have both a resonance at 5GHz, and are localized to the vortex lines in the case of the Hopfion, and the Bloch points in the case of the toron. It is worth noting, that the defects i.e., the Bloch points or vortex lines, play an important role, as they exhibit a more complicated microstructure than the surrounding bulk, allowing for a wider range of frequencies to exist.

**Impact of damping**

Further insight into the field-driven dynamics of Hopfions and torons is obtained by considering the impact of damping (Fig. 4). The complex spin wave resonances spectrum is only present for a low damping parameter. For a value of $\alpha$ = 0.001 there are nine peaks of varying intensity in the resonance spectrum (Fig. 4). in contrast, with a large damping parameter $\alpha$ =0.1, the resonant peaks broaden and merge into a single wide band around 4GHz with reduced amplitude.

The Hopfion mode h.2, which has the largest amplitude in the power spectral density is also most robust to damping. It is localized around the vortex lines at the top and bottom of the Hopfion, with an additional component in a ring in the center of the Hopfion. The second most robust mode with regard to damping, h.3. is also localized at the vortex line. This topological vortex line not only concentrates the dynamics of the Hopfions, it also increases their robustness under damping.

In the case of the toron the most robust mode, t.2., is localized not only around the Bloch points at the top and the bottom, but also couples to the edge of the disk. The toron's Bloch points are less excitable by an out of plane field than the Hopfion's vortex lines. The modes that are localized only around the Bloch points quickly disappear under damping, while the mode t.2., which is coupled to the edge of the disk, persists.

**Spin wave mode excitations under sinusoidal magnetic field**

In the highly damped regime spin wave dynamics can still be excited with sinusoidal applied fields, which produces a breathing response to an out of plane field. These responses follow the driving force after a period of transience and die out quickly after the sinusoidal field is removed.

To gain insight into the high damping breathing modes, we correlate the excitation in $m_z$ across different radii of the disk. A response function is computed for each timestep by integrating the fluctuation in $m_z$ at each radius individually,

$$\delta m_z(\rho, t) = \int (m_z(r,t) - m_z(r,0))\rho d\theta dz$$

where $\rho$ is the radial coordinate and $r = (\rho, \vartheta, z)$ is the position vector in cylindrical coordinates. Taking the product of the response functions for two different radii and integrating over time we construct the correlation matrices shown in Fig. 5a, b.

$$\delta m_z(\rho_j, \rho_k) = \int (\delta m_z(\rho_j, t)\delta m_z(\rho_k, t))dt.$$

The bright regions indicate pairs of radii in where $m_z$ increases or decreases concurrently, the dark regions indicate pairs of radii where as $m_z$ decreases at one radius as it increases at another.

For comparison, this correlation matrix is constructed for the Hopfion as well as the related target Skyrmion. The Hopfion response is more localized to the torus, while the target Skyrmion correlation is more spread out over the disk Fig. 5a, b. This results from the three-dimensional nature of the Hopfion. The Hopfion is localized to the torus, essentially a one-dimensional closed string. The target Skyrmion on the other hand is spread out over the entire disk.

The characteristic features in the dynamical behavior shown here could prove useful to differentiate experimentally between a Hopfion and a target Skyrmion texture. Figures 5c and 5d show simulated soft x-ray transmission microscopy data, which are only probing the magnetization projection along the z-axis, and cannot distinguish those two spin textures [17], which are different as shown in Fig. 5e, f. The distinct correlation matrices of the spin dynamics of these two spin textures would allow the confirmation of three-dimensional character of the spin textures.

**Conclusion**

We have found strong dependences of the resonant spin wave dynamics in 3D spin textures on their topology. Topological defects such as vortex lines in the case of Hopfions and Bloch points in the case of torons were identified as attractors to localized spin waves in these systems.

The collapse of vortex lines in Hopfions into Bloch points in torons upon applying external magnetic fields is accompanied by a sharp discontinuity in the spin wave spectrum. Though the modes are discontinuous in frequency space, in real space the spin waves are localized to analogous regions with topological defects (Bloch points and vortex rings) and large Heisenberg energy. These regions serve to focus the spin waves and provide a larger spectrum of excitable frequencies.

The correlated motions of these spin textures are also very specific. The resonances of a Hopfion are more localized at the torus than those of a target Skyrmion. The radial breathing modes of the target Skyrmion were found to hybridize with modes traveling vertically [12]. Hopfion spin waves are stronger localized but are in general not coupled to the boundary. Therefore, since the Hopfion is shielded by the uniformly magnetized background it is embedded in, the spin wave resonances are more stable against external perturbations, which could open an avenue to harnessing excitations in Hopfion lattices and their breathing mode as information carriers in three-dimensional architectures [29].

**Acknowledgements**

This work was funded by the U.S. Department of Energy, Office of Science, Office of Basic Energy Sciences, Materials Sciences and Engineering Division under Contract No. DE-AC02-05-CH11231 (Non-equilibrium magnetic materials program MSMAG).

This research used the Lawrencium computational cluster resource provided by the IT Division at the Lawrence Berkeley National Laboratory, supported by the U.S. Department of Energy, Office of Science, Office of Basic Energy Sciences under Contract No. DE-AC02-05CH11231)

**Figures**

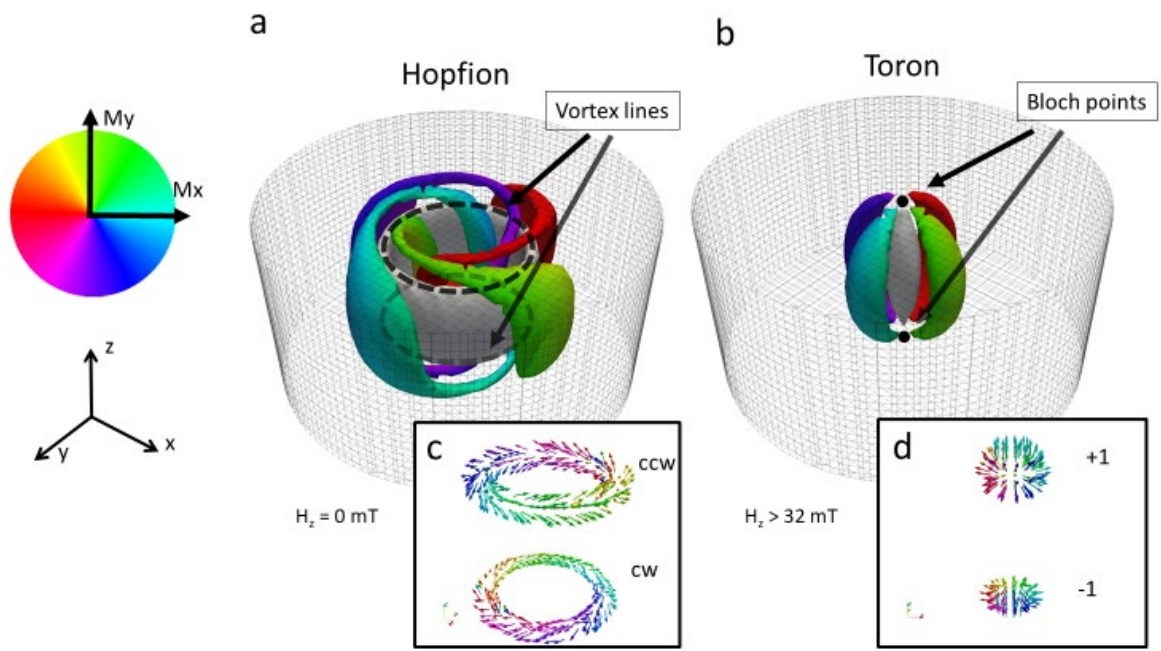

**Fig.1. (a)** Hopfion spin textures with linked magnetization iso-surfaces **(b)** toron spin textures with un-linked iso-surfaces **(c)** detail of Hopfion vortex rings **(d)** detail of toron Bloch points

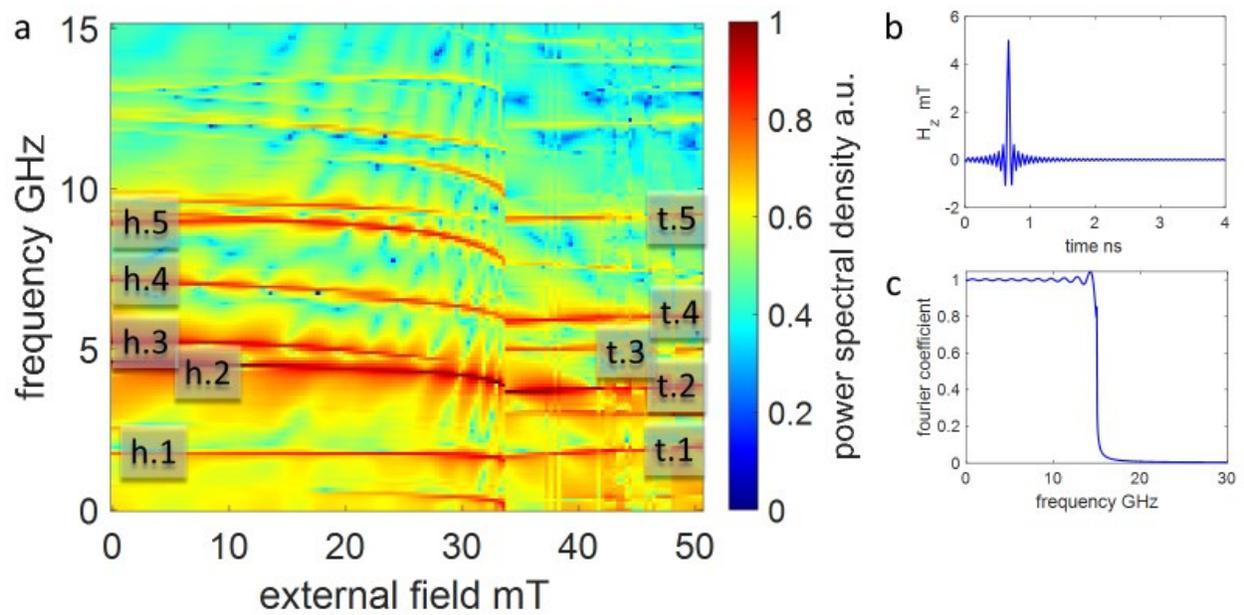

**Fig.2. (a)** Power Spectral Density (PSD) response of a Hopfion to a sinc pulse under static applied external field. The five lowest Hopfion and toron modes are labelled by h.1 – h.5 and t.1 – t.5, respectively. **(b)** Magnetic excitation pulse applied in the x-direction in the time domain. **(c)** Fourier transform of the excitation pulse in frequency domain.

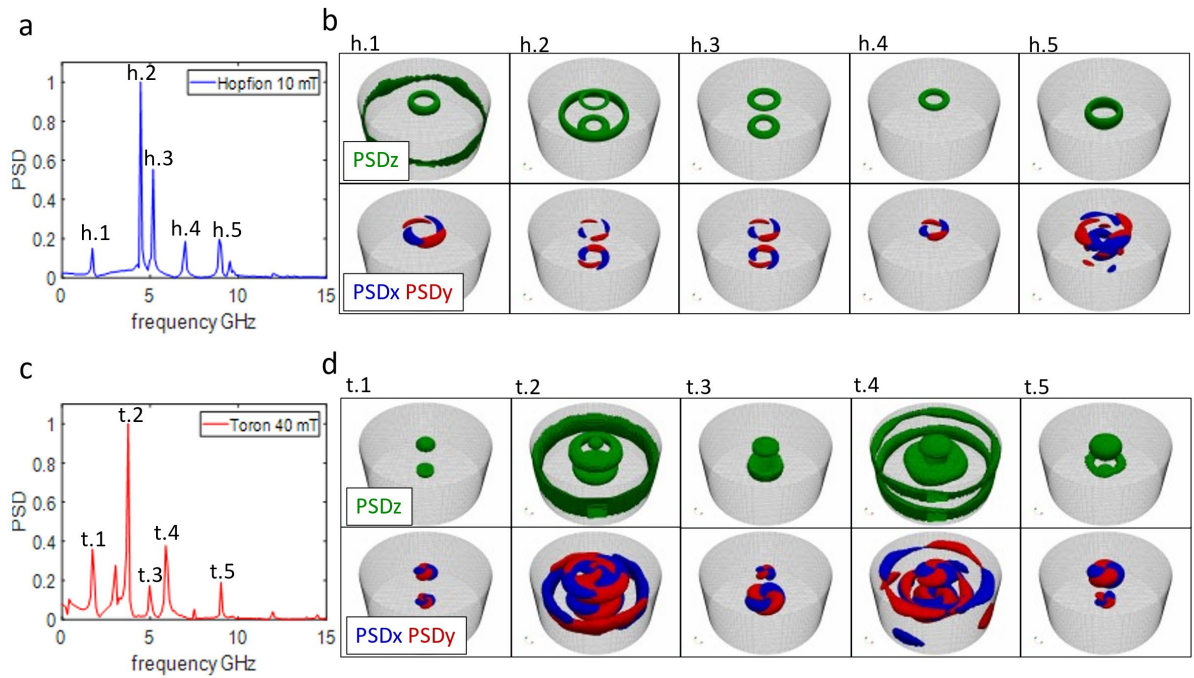

**Fig.3. (a)** Hopfion Power Spectral Density (PSD) at 10 mT. **(b)** Corresponding Hopfion spin wave modes h.1 - h.5 in real space. **(c)** Toron Power Spectral Density (PSD) at 40 mT. **(d)** Corresponding toron spin wave modes t.1- t.5 in real space. x,y,z coordinates are identical as in Fig 1.

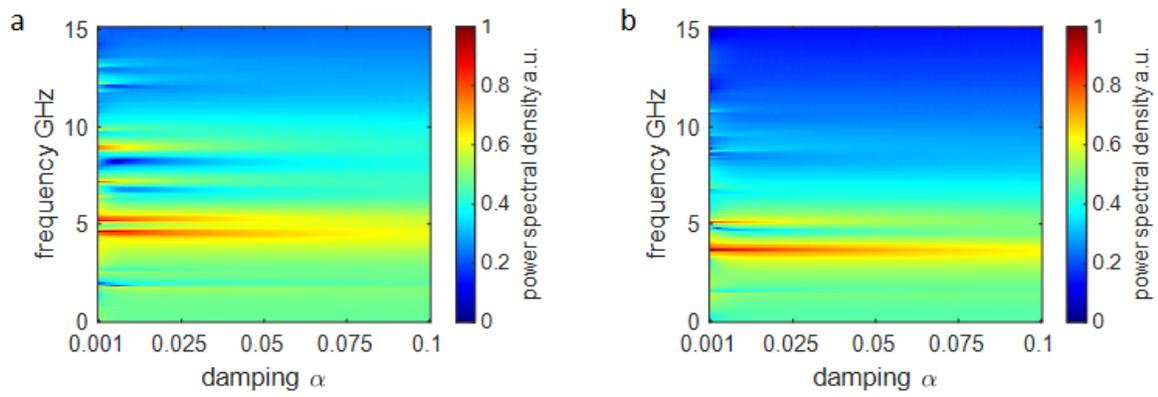

**Fig.4. (a)** Hopfion Power Spectral Density (PSD) at 10 mT under varying damping. **(b)** Toron Power Spectral Density (PSD) at 40 mT under varying damping.

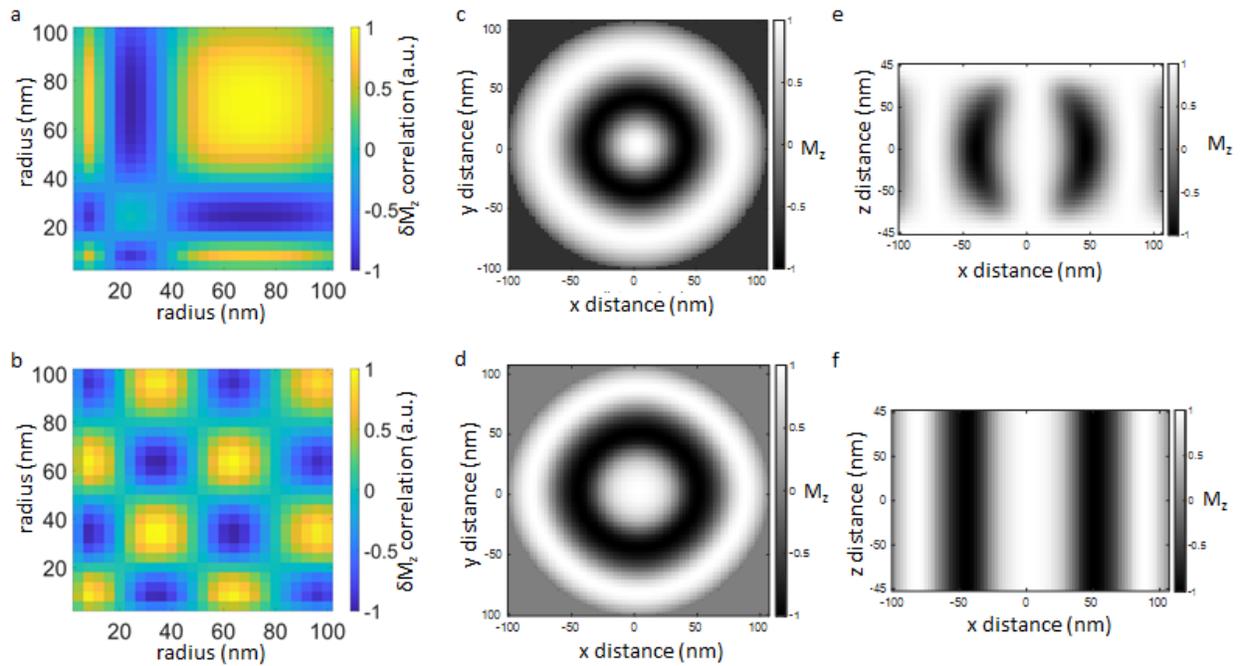

**Fig.5. (a)** Hopfion correlation matrix. **(b)** Target Skyrmion correlation matrix. **(c)** Hopfion $M_z$ magnetization at z=0 plane. **(d)** Target Skyrmion $M_z$ magnetization at z=0 plane. **(e)** Hopfion $M_z$ magnetization at y=0 plane. **(f)** Target Skyrmion $M_z$ magnetization at y=0 plane.